\documentstyle[prd,aps]{revtex}
\def \a {\alpha}
\def \b {\beta}
\def \g {\gamma}

\def \d {\delta}

\def \e {\eta}
\def \f {\phi}

\def \th {\theta}

\def \k {\kappa}
\def \l {\lambda}
\def \m {\mu}
\def \n {\nu}
\def \x {\xi}
\def \p {\pi}
\def \r {\rho}

\def \o {\omega}

\def \L {\pounds}

\def \der {\partial }
\def \nn {\nonumber}
\def \rov {\equiv}

\def \F {{\cal F}}

\def \K {{\cal K}}

\def \L {{\cal L}}

\def \J {{\cal J}}
\def \R {{\cal R}}

\def \U  {{\cal U}}
\def \T  {{\cal T}}
\def \sn {\sin \th}
\def \cs {\cos \th}
\def \tg {\tan \th}
\def \ctg {\cot \th}

\def \ct {\bar t}
\def \dsn {\sin^2 \th}

\def \tsn {\sin^3 \th}

\def \dctg {\cot^2 \th}
\def \chd  {\cosh 2\d}
\def  \shd  {\sinh 2\d}
\def \mm {\mbox{\quad }}

\def \msip {\rightarrow}

\def \lvkz {\Bigl(}
\def \pvkz {\Bigr)}

\def \lvhz {\Bigl[}
\def \pvhz {\Bigr]}
\def \lvvhz {\biggl[}
\def \pvvhz {\biggr]}

\def \lvvsz {\Biggl\{}
\def \pvvsz {\Biggr\}}
\def \pul {{{\scriptstyle{\frac{1}{2}}}}}

\def \ctvrt {{{\scriptstyle{\frac{1}{4}}}}}

\def \V  {\frac{V}{r}}

\def \cc {(c,_\th+2c\ctg)}
\def \dd {(d,_\th+2d\ctg)}

\def \VUW {\lvkz\V\edb-r^2\edg U^2\chd
           -r^2\emdg W^2\chd-2r^2 UW\shd\pvkz}

\def \edb  {{\rm e}^{2\b }}

\def \edg  {{\rm e}^{2\g }}
\def \emdg {{\rm e}^{-2\g }}
\def \eg   {{\rm e}^{\g }}
\def \emg  {{\rm e}^{-\g }}

\def \BE {\begin{equation}}
\def \EE {\end{equation}}
\def \BDM {\begin{displaymath}}
\def \EDM {\end{displaymath}}
\def \BEAH {\begin{eqnarray*}}
\def \EEAH {\end{eqnarray*}}
\def \BEA {\begin{eqnarray}}
\def \EEA {\end{eqnarray}}
\def \BM {\begin{math}}
\def \EM {\end{math}}
\def \mm {\mbox{\quad }}

\def \msip {\rightarrow}

\def \gab {g_{\a \b }}

\def \gjj {g_{11}}
\def \gjd {g_{12}}
\def \gjt {g_{13}}

\def \gdd {g_{22}}

\def \fk  {f^{(k)}}
\def \fmj {f^{(-1)}}
\def \fn  {f^{(0)}}
\def \fj  {f^{(1)}}

\def \Ak  {A^{(k)}}
\def \Amj {A^{(-1)}}
\def \An  {A^{(0)}}
\def \Aj  {A^{(1)}}

\def \gk  {g^{(k)}}
\def \gmj {g^{(-1)}}
\def \gen  {g^{(0)}}
\def \gej  {g^{(1)}}

\def \Tmn {T_{\m\n}}

\def \Tnn {T_{00}}
\def \Tnj {T_{01}}
\def \Tnd {T_{02}}
\def \Tnt {T_{03}}

\def \Tjj {T_{11}}
\def \Tjd {T_{12}}
\def \Tjt {T_{13}}

\def \Tdd {T_{22}}
\def \Tdt {T_{23}}

\def \Ttt {T_{33}}
\begin{document}
\draft
\title{{{\mbox{Symmetries} of \mbox{asymptotically} flat  
axisymmetric \mbox{spacetimes} with null dust}}}
\author{U. von der G\"{o}nna\footnote{Electronic mail:
{\tt U.v.d.Goenna@tpi.uni-jena.de}}\\}
\address{Institute for Theoretical Physics, FSU Jena,\protect\\
Max-Wien-Platz 1,\protect\\
D-07743 Jena, Germany}
\author{A.~Pravdov\' a\footnote{Electronic mail: 
{\tt pravdova@mbox.troja.mff.cuni.cz} and {\tt pravdova@math.cas.cz}}\\}
\address{Mathematical Institute, Academy of Sciences, \protect\\
\v Zitn\' a 25, \protect\\
115 67 Prague 1, Czech Republic}
\date{\today}
\maketitle
\begin{abstract}

Symmetries of spacetimes with null dust field as a source
compatible with asymptotic flatness are studied 
by using the~Bondi-Sachs-van der Burg formalism. 
It is  shown that in an axially symmetric spacetime
with null dust field in which
at least locally a smooth null infinity in the~sense
of Penrose exists,
the~only allowable additional Killing vector 
forming with the~axial one a two-dimensional Lie algebra
(the~axial and the~additional Killing vector are not assumed to
be hypersurface orthogonal)
%of axially symmetric spacetimes with null dust
 is a supertranslational Killing vector
%does not admit any other additional Killing vector
and the~gravitational field is then non-radiative 
(the~Weyl tensor has a non-radiative character).
\end{abstract}

%--------------------------------------

\section{Introduction and Summary}
%--------------------------------------

Recently a unique role of  boost-rotation symmetric
electrovacuum spacetimes describing 
``uniformly accelerated particles'' of
various kinds \cite{Gbicakobecne}
was  exhibited by a theorem which states
that in axially symmetric, asymptotically flat spacetimes
the~only additional
symmetry that does not exclude radiation is the~boost 
symmetry (in Ref.~\cite{Gbicak} for vacuum spacetimes with
hypersurface orthogonal Killing vectors and in
Ref.~\cite{Gajajibi} for electrovacuum spacetimes with
Killing vectors which are in general not hypersurface orthogonal).
Our effort in this paper is to prove a similar theorem
for asymptotically flat spacetimes with null dust fields. 
We also specialize the~spacetime to be axially symmetric 
(with the~axial Killing vector which is in general
not hypersurface orthogonal) - this
assumption simplifies lengthy calculations.

If one is interested in  gravitational radiation 
from a general bounded matter source, i.e., in  the~behaviour
of  gravitational field far from the~source, one has to turn
to approximation methods. 
The Bondi-Sachs formalism  is a powerful instrument for the treatment
not only of asymptotically flat vacuum \cite{Gbondi,Gsachs} 
and electrovacuum \cite{Gburg} fields, 
but also for  the investigation of asymptotically flat
null dust fields \cite{Ggonna}. 
The energy-momentum tensor of null dust or pure radiation
\BE
T_{\a\b}=\rho n_\a n_\b, \qquad n_\a n^\a=0,\qquad \rho>0\ , \label{Gtensor}
\EE
with $\r$ being the energy density
of the radiation field,
describes a field of massless radiation 
propagating along a null congruence
with the tangent vector $n^\a$. 
This field is the incoherent superposition of
waves with random phases and different polarizations 
where the radiation arises
from electromagnetic null field, massless scalar field,
neutrino field or gravitational field itself. 
The field equations of these originating fields are not considered.
Exact solutions of this class are the Vaidya solution 
\cite{Vaidya53} which
can model the exterior of a spherically symmetric shining star and 
the~Kinnersley photon rocket \cite{Kinn69}, 
a particle emitting photons 
and accelerating because of the recoil.
 Also null dust fields with rotation are
known \cite{Rot1,Rot2}.

In Sec.~II we start out from the~general form
of an axially symmetric metric 
in Bondi-Sachs coordinates $\{u,\ r,\ \th,\ \f\}$, where 
the~null coordinate $u$ and the~spherical angles 
$\th$, $\f$ are constant
along null rays while the~luminosity distance  $r$ varies.
We consider their asymptotic series
expansions at $r\ \to \infty$ 
assuming the~Einstein equations for the null dust field 
to be satisfied and the~spacetime to be asymptotically flat.

Then we assume that, in addition, another symmetry
exists, i.e.,  an additional Killing vector field
which forms with the~axial one a two-dimensional Lie algebra.
By first decomposing the~additional Killing
vector field $\e^\a$ in the~null  tetrad and then solving
the~Killing equations asymptotically in the~leading order
for this new Killing vector, we 
find that the~additional Killing vector asymptotically
generates either a supertranslation or a boost along the~symmetry axis.
However, developing and solving Killing equations in higher
orders and considering Lie derivatives of the~energy momentum
tensor for null dust to vanish for both
the~supertranslational Killing vector in Sec.~III and
the~boost Killing vector in Sec.~IV we conclude that in fact
the only allowable additional Killing vector of axially 
symmetric spacetimes with null dust is a supertranslational
Killing vector and then the~gravitational field is non-radiative 
(the~Weyl tensor has a non-radiative character).
 
Our conventions for the~Riemann and Ricci tensors follow those
of Ref. \cite{GMTW} but our signature is $-2$.

\section{{{\mbox{Axisymmetric} null dust \mbox{spacetimes}   
with \mbox{another} \mbox{symmetry}}}}

Consider an axially symmetric spacetime 
with the~corresponding Killing vector
field denoted by $\der / \der\f$. 
%filled with null dust. 
Assume that at least the~"piece of
$\J^+$" exists in the~sense of Ref.~\cite{AstSchm}. Then one can
introduce the~Bondi-Sachs coordinate system
\mbox{ \{ $u$,~$r$,~$\th$,~$\f$ \} $\rov$ \{
             $x^0$,~$x^1$,~$x^2$,~$x^3$\}}
in which the~metric %satisfying the~Einstein equations
has the~form \cite{Gbondi,Gsachs,Gburg}
\BEA
{\rm d}s^2&=&\VUW {\rm d}u^2\nn\\
    & &\ + 2\edb {\rm d}u {\rm d}r
              +2r^2(\edg U\chd +W\shd)  {\rm d}u{\rm d}\th
              +2r^2(\emdg W\chd +U\shd)\sn\ {\rm d}u{\rm d}\f\nn\\
    & &\ - r^2\left[ \chd(\edg {\rm d}\th^2 +\emdg \dsn\ {\rm d}\f^2 )
                        +2\shd\sn\ {\rm d}\th {\rm d}\f\right] \ ,\label{ds}
\EEA
where the~six metric functions $U$, $V$, $W$, $\b$, $\g$, $\d$ 
%and the~null dust field described by 
%an energy-momentum tensor $T_{\m\n}$ 
do not depend on $\f$ because of axial
symmetry.
The spacetime (\ref{ds}) is filled with null dust described 
by the~energy-momentum tensor (\ref{Gtensor}) 
which is assumed to be axially symmetric, too. 

In Ref.~\cite{Ggonna} the~expansions of the~metric 
functions of (\ref{ds}) 
for asymptotically flat spacetimes with null dust 
in Bondi-Sachs coordinates are derived.
For that, the~null expansion vector $n^\a$ is chosen to
be identified with the~null vector $k^\a$ of 
the~Bondi-Sachs tetrad (\ref{Btetrada}) at ${\cal J}^+$,
i.e., ${\cal J}^+$ has to exist in the direction determined by
the null vector $n^\a$ of the null dust. 
Then its contravariant components are
\BEA
n^u   &=&  \frac{\U (u,\ \th)}{r^2}+{\cal O} (r^{-3})  \ ,\nn\\
n^r   &=&1+\frac{\R (u,\ \th)}{r}+{\cal O} (r^{-2}) \ ,\nn\\
n^\th &=&  \frac{\T (u,\ \th)}{r^2}+{\cal O} (r^{-3}) \ ,\label{Gnulvektorkontr}\\
n^\f  &=&  \frac{\F (u,\ \th)}{r^2}+{\cal O} (r^{-3}) \ ,\nn
\EEA
and the~covariant components read 
\BEA
n_u   &=&1+\frac{\R}{r}+{\cal O} (r^{-2}) \ ,\nn\\
n_r   &=&  \frac{\U }{r^2}+{\cal O} (r^{-3}) \ ,\nn\\
n_\th &=&-\T+{\cal O} (r^{-1}) \ ,\label{Gnulvektorkov}\\
n_\f  &=&-\F \ \dsn+{\cal O} (r^{-1}) \ .\nn
\EEA
As the~vector $n^\a$ is null ($n^\a n_\a=0$),
functions entering (\ref{Gnulvektorkontr}), 
(\ref{Gnulvektorkov}) have to satisfy
\BE
\U =\pul ({\T}^2+{\F }^2\dsn)\ ,\label{Gpodmnulvektor}
\EE
and similarly for the~higher-order coefficients. 
Then the equations for the~null dust field 
(\ref{Gtensor}) can be solved.
The~metric coefficients have in the~first order in $r^{-k}$ the~same
form as Eq.~(4) in Ref.~\cite{Gajajibi}:
\BEA   
\g  &=&\frac{c}{r}+{\cal O}(r^{-3})\ ,\nn\\
\d  &=&\frac{d}{r}+{\cal O}(r^{-3})\ ,\nn\\
\b  &= & -\frac{1}{4}(c^2+d^2)\frac{1}{r^2}+{\cal O}(r^{-4})\ ,\nn\\
U   &= &-\cc\frac{1}{r^2}+{\cal O}(r^{-3})\ ,\label{rozvmetr}\\
W   &= &-\dd\frac{1}{r^2}+{\cal O}(r^{-3})\ ,\nn\\
V   &= &r-2M+{\cal O}(r^{-1})\ .\nn
\EEA
For  the radiation density  $\r(u,\ r,\ \th)$ we write
\BE
\r(u,\ r,\ \th)=\frac{\r_2(u,\ \th)}{r^2}+{\cal O}(r^{-3})\ \label{Ghustota}
\EE
and from the field equations 
\BE
M,_u=-(c,_u^2+d,_u^2)-\pul \k_0\r_2
     +\pul (c,_{\th\th}+3c,_\th\ctg-2c),_u\label{GMu}
\EE
follows. The~energy balance at null infinity 
(where null infinity admits a regular spherical cross section)
then shows that the mass loss $m,_u$
results from a linear superposition of the pure and 
the gravitational radiation 
parts \cite{Ggonna}
\BE
m,_u=-\pul\int\limits_{0}^{\p} (c,_u^2+d,_u^2+\pul \k_0\r_2)\sn {\rm d}\th
               \leq 0\,\label{balance}
\EE
with the~function 
$\r_2(u,\ \th)$ being an analogue to the~news functions
of electromagnetic field squared $X^2+Y^2$
(see  (14) in \cite{Gburg}).

Since we admit spacetimes with only "local" $\J^+$, we assume 
Eqs.~(\ref{ds})--(\ref{balance}) to be satisfied 
for $\f\in\langle 0,2\p)$, however 
not necessarily on the~whole sphere, i.e., 
for all $\th\in\langle 0,\p\rangle $, 
but only in some open interval of $\th$. 

Let us follow a similar procedure to that one used
for the~electrovacuum case and assume here again 
the~spacetime to have another Killing vector $\e$ which 
forms a two-dimensional Lie algebra with the~axial one,
$\x =\der / \der\f$,
i.e., we  assume $[\e,\x]=0$ 
(see the~Lemma in Sec. 2 in \cite{Gbicak}). 
Hence, the~components of $\e^\a$ are independent of $\f$.

We introduce the~standard null tetrad 
$\lbrace k^{\a }, m^{\a }, t^{\a }, \ct ^{\a }\rbrace$
(for details see (11) and the~paragraph above
in \cite{Gajajibi}), with bar
denoting the~complex conjugation:
\BEA
k_\a&=&[1\ ,\ 0\ ,\ 0\ ,\ 0]\ ,\mm
m_\a=[\pul Vr^{-1}\edb\ ,\ \edb\ ,\ 0\ ,\ 0]\ ,\nn\\
t_\a&=&\pul r(\chd)^{-\pul}
        \lvhz (1+\shd)\eg U+\chd\emg W
           +{\rm i}[(1-\shd)\eg U-\chd\emg W]\ ,\label{Btetrada}\\
    &\ &\mm 0\ ,\
         -(1+\shd+{\rm i}(1-\shd))\eg\ ,\ -(1-{\rm i})\chd\sn\emg\pvhz\ ,\nn
\EEA
and decompose the~additional
Killing vector $\e^\a$ in this null tetrad 
\BE
\e ^\a  =Ak^\a +Bm^\a +\tilde{f}(t_{R}^{\a}+t_{I}^{\a})
                      +\tilde{g}(t_{R}^{\a}-t_{I}^{\a})\ ,\label{GBeta}
\EE
where $A$, $B$, $\tilde{f}$, $\tilde{g}$ 
are general functions of $u$, $r$, $\th$ and $t^\a=t_R^\a+{\rm i}t_I^\a$ .

The~Killing vector $\e^\a$ has to satisfy the~Killing equations
(all of them are written down in \cite{Gaja})
\BE
\L_\e \gab=0\ .\label{Kill}
\EE
The~easiest one among them is the~equation
\BE
\L_\e \gjj=2\edb B,_r=0 \ ,\nn
\EE
which implies
\BE
B=B(u,\th)\ .\label{B}
\EE

We solve the~other Killing equations
asymptotically assuming that the~coefficients 
$A$, $\tilde{f}$, $\tilde{g}$  can be expanded in
powers of $r^{-k}$. 
Then equations $\L_\e \gdd=0$, \mbox{$\L_\e \gjd=0$}, 
$\L_\e \gjt=0$ imply
\BEA
A&=&\Amj r+\An+\frac{\Aj}{r}+{\cal O}(r^{-2})\ ,\nn\\
\tilde{f}&=&\fmj r+\fn+\frac{\fj}{r}+{\cal O}(r^{-2})\ ,\label{Afg}\\
\tilde{g}&=&\gmj r+\gen+\frac{\gej}{r}+{\cal O}(r^{-2})\ ,\nn
\EEA
where the~coefficients 
$\Ak$, $\fk$, $\gk$ are functions of $u$ and $\th$.

Since the~null dust field decays at infinity 
in the~same way as the~electromagnetic field
in \cite{Gajajibi},
it does not enter the~Killing equations in 
the~leading order in $r^{-k}$ as in \cite{Gajajibi} 
(see Eqs. (19)--(25) therein)
and their solution is thus identical to the~solution
in the~electrovacuum case in \cite{Gajajibi}
and even the~solution obtained in the~vacuum case 
examined in \cite{Gbicak}:
\BEA
\Amj&=&k\cs\ ,\nn\\
\fmj&=&-k\sn\ ,\label{BAmjfmjB}\\
B   &=&-ku\cs+\a(\th)\ ,\nn
\EEA
where $k=\mbox{const}$ and $\a$ is an arbitrary function of $\th$
and 
\BE
\gmj=h\sn\ ,\label{Bgmj}
\EE
where $h=\mbox{const}$.
One can easily find (using Eqs. (\ref{Btetrada}), 
(\ref{GBeta}) and (\ref{Afg})) that 
the~contribution of $h$ to the~vector
field $\e^\a$ is just constant multiple of the~axial 
Killing vector  $\der /\der\f$, $\e^\f=h+{\cal O}(r^{-1})$,
and so we may,
without loss of generality, put $h=0$.
Therefore, in the~lowest order of $r^{-1}$ the~general
asymptotic form of the~Killing vector $\e$ turns out to
be
\BE
\e^\a=[-ku\cs+\a(\th)\ ,\
          kr\cs+{\cal O}(r^{0})\ ,\ -k\sn+{\cal O}(r^{-1})\ ,\ 
          {\cal O}(r^{-1})]\ , 
\label{GBbotr}
\EE
where $k$ is a constant, $\a$ -- an arbitrary function of $\th$.
Thus, 
assuming the~presence of a null dust field satisfying
the~boundary conditions (\ref{Gnulvektorkontr})--(\ref{Gpodmnulvektor})
and Killing vectors which need not be hypersurface orthogonal,
we arrive
in the~leading order of the~asymptotic expansion 
at the~same conclusion obtained
in Ref. \cite{Gbicak} 
in the~vacuum case with hypersurface orthogonal $\der /\der\f$
or in Ref.~\cite{Gajajibi} for the~electrovacuum case which
Killing vectors are not hypersurface orthogonal.
When $k=0$, the~vector field
(\ref{GBbotr}) generates supertranslations.

Assuming $k\not= 0$,
one can find a Bondi-Sachs coordinate system with $\a=0$ by making
a supertranslation, as was shown in \cite{Gajajibi}.
Hence, we put $\a=0$ in Eq. (\ref{GBbotr}) and without loss
of  generality  we choose $k=1$. Then $B=-u\cs$,
$\Amj=\cs$, $\fmj=-\sn$ and $\gmj=0$
and the asymptotic form of the~Killing vector field
$\e$ is
\BE
\e^\a=[-u\cs\ ,\ r\cs+{\cal O}(r^{0})\ ,
\ -\sn+{\cal O}(r^{-1})\ ,\ {\cal O}(r^{-1})]\ ,\label{vboost}
\EE
that is the~boost Killing vector. It  generates
the~Lorentz transformations along the~axis of axial symmetry.

The~conclusion of this section is thus following:\\
Suppose that an axially symmetric
spacetime with null dust admits a ``piece'' of $\J^+$ in the~sense
that the~Bondi-Sachs coordinates can be introduced in which
the~metric takes the~form (\ref{ds}), (\ref{rozvmetr}) 
and the~asymptotic forms of the~energy-mass 
density and the null vector 
field of the~null dust is given by
(\ref{Gnulvektorkontr})--(\ref{Gpodmnulvektor}). 
If this spacetime admits an additional Killing
vector forming with the~axial Killing vector a two-dimensional Lie
algebra, then the~additional Killing vector has asymptotically
the~form (\ref{GBbotr}). For $k=0$ it generates a
supertranslation; for $k\not= 0$ it 
generates a boost along the~symmetry axis.

However in the~next sections we see that 
the~boost symmetry is in fact not allowable.
\section{{{\mbox{The~supertranslational} 
\mbox{Killing} \mbox{field}}}}
%--------------------------------

In this section, assuming $k=0$ for 
the~Killing field (\ref{GBbotr}),
we consider the~Killing equations in higher orders 
of $r^{-1}$ and arrive at the~same equations 
as (31)--(48) in \cite{Gajajibi} %:
with the~same solutions (49)--(55) therein: 
\BEA
c&=&\pul uB^{-1}(B,_{\th\th}-B,_\th\ctg)\ ,\mm
d=d(\th)\ ,\mm M=-uc,_u^2-B^{-1}[\Aj+B,_\th\cc]\ ,\nn\\
\An &=&\pul(B,_{\th\th}+B,_\th\ctg+B)\ ,\mm
\fn=\fn(\th)=-B,_\th\ ,\mm 
\gen=0\ , %\label{GTAfgnula}\\
\label{GTcd}\\
\Aj &=&\Aj(\th)\ ,\mm
\fj =B\cc\ ,\mm
\gej=B\dd-B,_\th d\ ,\nn %\label{GTAfgjedna}
\EEA
where $B$, $\Aj$ and $d$ are arbitrary functions of $\th$.
Thus, in the~supertranslational case for the~null dust 
the~Weyl tensor is also non-radiative as it was in electrovacuum
spacetimes.
Substituting the~metric functions (\ref{rozvmetr})  
into the~null tetrad (\ref{Btetrada}) 
and coefficients (\ref{GTcd}), the~expansion of 
the~Killing vector %$\e$ 
reads
\BEA
\e^\m&=&\lvhz B(\th)\ ,\ \pul (B,_{\th\th}+B,_\th\ctg)+
               [-B,_{\th\th}-2B,_\th B,_{\th\th\th}
             +2{B,_\th}^2B,_{\th\th}B^{-1} -2{B,_\th}^3\ctg B^{-1}
     \nn\\      &\ &\mm 
           +{B,_\th}^2 (3\ctg^2-2\sin^{-2}\th)]B^{-1}\frac{u}{4r}
                    +{\cal O}(r^{-2})\ ,  
% \nn\\& \ &
       \ -B,_\th \frac{1}{r}+B,_\th\frac{c}{r^2}+{\cal O}(r^{-3})\ ,
       \  B,_\th \frac{d}{r^2\sn}+{\cal O}(r^{-3})\pvhz\ .
            \label{Getasupertrans}
\EEA

Now let us turn to the~asymptotic behaviour of the~null dust.
It is easy to show that if $\e$ is a Killing vector then 
the~Lie derivative of the~Riemann tensor with respect to 
this vector vanishes, and then also the~Lie derivative of the~Ricci
tensor vanishes. And if, in addition, the~Ricci scalar is zero
and Einstein's equations are satisfied, 
then the~following equations hold
\BE
\L_\e T_{\m\n}=0\ .\label{GLieTmn}
\EE
Substituting $\e$ from (\ref{Getasupertrans}), equations (\ref{GLieTmn})
get in the~leading order the~form 
\BEA
L_\e \Tnn=0\mm (r^{- 2}):&& \mm  \r_2,_u B=0\ ,\label{GTLienn}\\
L_\e \Tnj=0\mm (r^{- 4}):&& \mm  \r_2,_u \U  B+\ \r_2(\U ,_u B-\T B,_\th)=0\ ,
           \label{GTLienj}\\
L_\e \Tnd=0\mm (r^{- 2}):&& \mm  \r_2,_u \T B+\ \r_2(\T,_u B-B,_\th)=0\ ,
           \label{GTLiend}\\
L_\e \Tnt=0\mm (r^{- 2}):&& \mm  \r_2,_u \F  B+\ \r_2\ \F ,_u B=0\ ,
            \label{GTLient}\\
L_\e \Tjj=0\mm (r^{- 6}):&& \U [\r_2,_u \U  B+2 \r_2(\U ,_u B-\T B,_\th)]=0\ ,
            \label{GTLiejj}\\
L_\e \Tjd=0\mm (r^{- 4}):&& \mm  \r_2,_u \U  \T B
%   \nn\\ &&\mv
              +\r_2[(\U  \T),_u B-(\U  +{\T}^2)B,_\th]=0\ ,
            \label{GTLiejd}\\
L_\e \Tjt=0\mm (r^{- 4}):&& \mm  \r_2,_u \U  \F  B
              +\r_2[(\U \F ),_u B-\T \F  B,_\th]=0 ,
             \label{GTLiejt}\\
L_\e \Tdd=0\mm (r^{- 2}):&& \T[\r_2,_u \T B+2\r_2(\T,_u B-B,_\th)]=0\ ,
             \label{GTLiedd}\\
L_\e \Tdt=0\mm (r^{- 2}):&& \mm  \r_2,_u \T \F  B
              +\r_2[(\T \F ),_u B-\F  B,_\th]=0\ ,\label{GTLiedt}\\
L_\e \Ttt=0\mm (r^{- 2}):&& \mm  \r_2,_u \F  B+2\r_2 \F ,_u B=0\ .\label{GTLiett}
\EEA
If $\r_2=0$, then all equations are trivially satisfied and it can be 
shown that we deal with
a vacuum spacetime with an arbitrary supertranslational symmetry.
If we suppose $B$ and $\r_2$ to be non-vanishing, then 
the~first equation (\ref{GTLienn}) implies
\BE
\r_2=\r_2(\th)\ ,
\EE
and  from Eqs.~(\ref{GTLient}) or (\ref{GTLiett}), 
(\ref{GTLiend}) or (\ref{GTLiedd}), and (\ref{GTLienj}), (\ref{GTLiejj})
or (\ref{GTLiejt}) it  follows
\BEA
\F ,_u&=&0\ ,\\
\T,_u&=&B,_\th B^{-1}\ ,\\
\U ,_u&=&B,_\th B^{-1}\ \T\ ,
\EEA
which yield
\BEA
\F &=&\F_{0}(\th)\ ,\\
\T&=&B,_\th  B^{-1}\ u+\T_{0}(\th)\ ,\\
\U &=&\pul B,_\th^2B^{-2}\ u^2 +B,_\th B^{-1}\T_{0}u
      +\pul ({\T_{0}}^2+{\F_{0}}^2\dsn)\ ,
\EEA
where we used condition (\ref{Gpodmnulvektor}).
Comparing (\ref{GMu}) with (\ref{GTcd}) we obtain
the~following equation determining the~function $B(\th)$
for the~given ``null dust'' news function $\r_2(\th)$:
\BE
\lvvsz \frac{B^2}{\sn}
\lvvhz \frac{\tsn}{2B}\lvkz\frac{B,_\th}{\sn}\pvkz,_\th\pvvhz,_\th
\pvvsz,_\th =\k_0\r_2 B^2\sn\ .
\EE

If a spacetime admits a global null infinity this %This 
case with a non-zero $\r_2$ is not very physical since 
(\ref{balance}) then implies  permanent linear 
decreasing of the~total Bondi mass $m$.

\section{{{\mbox{The~boost} \mbox{Killing} \mbox{field}}}}
%---------------------------------------------------
%

In this section we investigate the~boost case, $k=1$ and
$\a=0$, similarly as the~previous one
by expanding the~Killing equations in higher orders of $r^{-1}$.
We obtain the~same conditions  for the~coefficients $\Ak$,
$\fk$ and $\gk$ as Eqs.~(81)--(89) and (104)--(112) 
in \cite{Gajajibi} %:
 with %having 
the~identical  solutions
\BEA
\An(u,\th)&=&\pul u\cs\ ,\label{GbAn}\\
\fn(u,\th)&=&-u^2w+\K(w)=-u\sn+\K(\sn /u)\ ,
%\mm\mbox{where}\ ??w=\frac{\sn}{u}\ ,
\label{Gbfn}\\
\gen(u,\th)&=&\gen(w)\ , \label{Gbgn}\\
\Aj(u,\th)&=&\frac{\cs}{w} \lvkz\pul\K,_{ww}w^2+2\K,_w w+\K\pvkz
    %  \nn\\  &&\mm     
    -\frac{\cs}{8\dsn}\lvkz 4\K^2+{\gen}^2\pvkz
   +\frac{\cs}{u^2}\L(w)
            \ ,\label{GbAj}\\
\fj(u,\th)&=&\frac{1}{4\sn}(-2\K^2+{\gen}^2)+u\dctg (\K,_w w+\K)
          \ ,\label{Gbfj}\\
\gej(u,\th)&=&\pul u\gen-\frac{\gen\K}{\sn}
           +\pul u\dctg(\gen,_w w+\gen)\ ,\label{Gbgj}\\
c(u,\th)&=& %-u-\frac{\fn}{\sn}=
      -\frac{\K(w)}{uw}\mm\msip\ 
     c,_u(u,\th)=\frac{\K(w),_w}{u^2}\ ,\label{Gbccu}\\
d(u,\th)&=&-\frac{\gen(w)}{2\sn}\mm\msip\ 
     d,_u(u,\th)=\frac{\gen(w),_w}{2u^2}\ ,\label{Gbddu}\\
M(u,\th)&=&\frac{1}{2u}(\K,_{ww}w+2\K,_w)+\frac{\L(w)}{u^3}
%    \nn\\  &=&
  =\frac{1}{2\sn}(w^2\K,_w),_w+\frac{\L}{u^3}\ ,\label{GbM}
\EEA
where  $\K(w)$ and $\gen(w)$   are arbitrary functions of $w$
\BE
 w=\frac{\sn}{u}\ ,\label{Gw}
\EE
and the~integration function $\L(w)$ entering the~expression
for the~mass aspect has the~form
\BE
\L(w)=\frac{\l(w)}{w^3}\ ,\label{GboostL}
\EE
with $\l$ satisfying the~equation
\BE
\l(w),_w=w^2(\K,_w^2+\ctvrt\gen,_w^2+\pul\k_0\o)
-\frac{1}{2w}(w^3\K,_{ww}),_w\ .\label{Gboostlambda}
\EE
Here, $\K$ and $\gen$ determine the~gravitational news functions,
$c,_u$ and $d,_u$, by the~relations (\ref{Gbccu}) and (\ref{Gbddu}),
$\o$ determines the~null dust news function,
$\r_2$, given below by the~relation (\ref{Gbhustota}).
Hence, solving the~last equation for $\l$ for  {\it{given}}
$\K$, $\gen$, and $\o$, we find $\L(w)$ and thus
the~mass aspect $M(u,\th)$ in the~form of Eq. (\ref{GbM}).
The~total mass $m$ at $\J^+$ is then given 
by integrating Eq. (\ref{GbM}) over the~sphere:
\BE
m(u)=\pul\int\limits_{0}^{\p} M(u,\th)\sn {\rm d}\th\
     =\ctvrt\int\limits_0^\p (w^2\K,_w),_w{\rm d}\th
     +\pul\int\limits_0^\p \frac{w\L}{u^2}{\rm d}\th\ .\label{Gboostm}
\EE

Substituting the~expansions of the~metric functions,
Eq. (\ref{rozvmetr}) % (resp. (4) in \cite{Gajajibi}),
into the~null tetrad, Eq. (\ref{Btetrada}), %(2.11) therein, 
and coefficients $A$, $B$, $\tilde{f}$ and $\tilde{g}$, 
Eqs. (\ref{GbAn})--(\ref{GbM}), into Eq.~(\ref{GBeta}), 
we find the~asymptotic form %expansion
of the~boost Killing vector to be
\BEA
\e^\m&= &
      \lvhz -u\cs\ ,\ r\cs+u\cs+\cs\ \lvkz\K,_w +\frac{\K}{w}\pvkz
                      \ \frac{1}{r}
              +{\cal O}(r^{-2})\ ,\ \nn\\
     &\ &\mm
             -\sn-u\sn\ \frac{1}{r} + uc\sn\ \frac{1}{r^2}+{\cal O}(r^{-3})\ ,
 %  \nn\\ &\ &
         \mm  \ ud\ \frac{1}{r^2}+{\cal O}(r^{-3})\pvhz\ .\label{Getaboost}
\EEA

Finally, let us turn our attention to the~asymptotic properties
of the~null dust represented by the~energy-momentum tensor $\Tmn$ 
which, as was shown in the~previous section, has to have the~vanishing 
Lie derivative (\ref{GLieTmn}) with respect
to the~Killing vector $\e^\a$, (\ref{Getaboost}).  
Regarding (\ref{GbAn})--(\ref{GbM}), these equations 
in the~first orders 
%substituting (\ref{GbAn})--(\ref{GbM}) 
look as follows:
\BEA
\L_\e \Tnn=0\mm (r^{- 2}):&& -\cs(u\r_2,_u+\tg\ \r_2,_\th-4\r_2)=0\ ,
          \label{GbLienn}\\
\L_\e \Tnj=0\mm (r^{- 4}):&& -\cs[(u\r_2,_u+\tg\ \r_2,_\th)\U 
            +\r_2(u\ \U ,_u+\tg\ \U ,_\th
%        \nn\\  &&\mm
          +4\U +u\tg\ \T)]=0\ ,
               \label{GbLienj}\\
\L_\e \Tnd=0\mm (r^{- 2}):&& \mm \cs[(u\r_2,_u+\tg\ \r_2,_\th)\T
            +\r_2(u\ \T,_u+\tg\ \T,_\th
%         \nn\\  &&\mm
         +4\T+u\tg)]=0\ ,
                \label{GbLiend}\\
\L_\e \Tnt=0\mm (r^{- 2}):&& \mm \cs[(u\r_2,_u+\tg\ \r_2,_\th)\F 
%     \nn\\  &&\mm
            +\r_2(u\ \F ,_u+\tg\ \F ,_\th+5\F )]=0\ ,
                \label{GbLient}\\
\L_\e \Tjj=0\mm (r^{- 6}):&& -\cs[(u\r_2,_u+\tg\ \r_2,_\th)\U 
            +2\r_2(u\ \U ,_u+\tg\ \U ,_\th
%     \nn\\   &&\mm
      +2\U +u\tg\ \T)]=0\ ,
              \label{GbLiejj}\\
\L_\e \Tjd=0\mm (r^{- 4}):&& \mm \cs\{ (u\r_2,_u+\tg\ \r_2,_\th)\U \T
            +\r_2[u(\U  \T),_u
    \nn\\      &&\mm
         +\tg\ (\U  \T),_\th
            +4\U \T
%         \nn\\  &&\mm
        +u\tg\ (\U +{\T}^2)]\}=0\ ,\label{GbLiejd}\\
\L_\e \Tjt=0\mm (r^{- 4}):&& \mm \cs\{ (u\r_2,_u+\tg\ \r_2,_\th)\U \F 
            +\r_2[u(\U  \F ),_u
%      \nn\\    &&\mm
        +\tg\ (\U  \F ),_\th
              +5\U \F +u\tg\ \T\F ]\}=0\ ,\label{GbLiejt}\\
\L_\e \Tdd=0\mm (r^{- 2}):&& -\cs[(u\r_2,_u+\tg\ \r_2,_\th)\T
%          \nn\\ &&\mm
            +2\r_2(u\ \T,_u+\tg\ \T,_\th+2\T+u\tg)]=0,\label{GbLiedd}\\
\L_\e \Tdt=0\mm (r^{- 2}):&& -\cs\{ (u\r_2,_u+\tg\ \r_2,_\th)\T\F 
            +\r_2[u(\T \F ),_u
%       \nn\\   &&\mm
            +\tg\ (\T \F ),_\th
             +5\T\F +u\tg\ \F ]\}=0\ ,\label{GbLiedt}\\
\L_\e \Ttt=0\mm (r^{- 2}):&& -\cs[(u\r_2,_u+\tg\ \r_2,_\th)\F 
%              \nn\\  &&\mm
            +2\r_2(u\ \F ,_u+\tg\ \F ,_\th+3\F )]=0\ .\label{GbLiett}
\EEA
The~trivial solution is again $\r_2=0$ which implies 
a vacuum boost-rotation symmetric spacetime. 
Let us assume $\r_2\not=0$.
Using variable $w$ given by (\ref{Gw}), 
Eq.~(\ref{GbLienn}) can be solved
to yield
\BE
\r_2=\frac{\o(w)}{u^2}\ ,\label{Gbhustota}
\EE
with an arbitrary function $\o(w)$. The~sum of Eqs.~(\ref{GbLient}) 
and  (\ref{GbLiett}) gives 
\BE
\F =\frac{\F_0(w)}{u}\ ,\label{GbNF}
\EE
($\F_0(w)$ being an arbitrary function) and
their difference is then identically zero. Similarly, summing
Eqs.~(\ref{GbLiend}) and  (\ref{GbLiedd}), we obtain the~equation
for $\T$,
\BE
\T,_u=-\tg=-\frac{uw}{\sqrt{1-u^2w^2}}\ ,
\EE
which leads to the~solution 
\BE
\T=\frac{\sqrt{1-u^2w^2}}{w}+\T_0(w)=u\ctg+\T_0(w)\ ,\label{GbNT}
\EE
where $\T_0(w)$ is an arbitrary integration function.
Next, the~difference of Eqs. (\ref{GbLiend}) and  (\ref{GbLiedd})
identically vanishes.
We repeat the~procedure for the~sum of Eqs.~(\ref{GbLienj}) and  
(\ref{GbLiejj}) and find the~solution
\BE
\U =-\pul u^2+\T_0(w)\frac{\sqrt{1-u^2w^2}}{w}+\U_0(w)
     =-\pul u^2+u\ctg\ \T_0(w)+\U_0(w)\ ,\label{GbNU}
\EE
($\U_0(w)$ is an arbitrary function of $w$). Then 
their difference is identically zero and  also 
Eqs.~(\ref{GbLiejd}), (\ref{GbLiejt}), (\ref{GbLiedt})  are identically
satisfied.

The~coefficients $\U $, $\T$ and $\F $ have, in addition, to fulfil
the~condition for the~null vector (\ref{Gpodmnulvektor}).
This, however, is in contradiction with 
the~boost-rotation symmetric solutions (\ref{GbNF}), (\ref{GbNT})
and (\ref{GbNU}).
Consequently, there are no asymptotically flat
boost-rotation symmetric solutions
of the~Einstein equations with null dust.
And the final conclusion reads:\\
{\bf Theorem}\\ 
Suppose that an axially symmetric
spacetime with null dust admits a ``piece'' of $\J^+$ in the~sense
that the~Bondi-Sachs coordinates can be introduced in which
the~metric takes the~form (\ref{ds}), (\ref{rozvmetr}) 
and the~asymptotic forms of the radiation 
density and the null vector 
field of the~null dust are given by
(\ref{Gnulvektorkontr})--(\ref{Gpodmnulvektor}). 
If this spacetime admits an additional Killing vector
forming a two-dimensional Lie
algebra with the~axial Killing vector, then the~additional
Killing vector which has asymptotically the~form 
(\ref{Getasupertrans}) generates asymptotically
supertranslations and the~Weyl tensor is non-radiative
(although one of the~``gravitational'' news functions, $c,_u$,
is non-vanishing, however, it is a function only of $\th$ as
the~``null dust'' news function, $\r_2$).

%----------------------------------------\\[2mm]
\begin{center}
{\bf{ACKNOWLEDGMENTS}}\\
\end{center}
We are grateful to  prof. {Bi\v c\' ak} for 
suggesting the topic and for discussions. 
%The work was supported by DFG-project
%KR 1219/4-2 and by the~grants nos. GACR-201/98/1452 
%and GACR-202/00/P031
%of the~Czech Republic. AP also acknowledges the hospitality of the Friedrich-Schiller-University in Jena.

\end{document}